\pdfoutput=1
\documentclass[aps,prl,amsmath,amssymb,twocolumn,twoside,showpacs]{revtex4-2}
\usepackage[pdftex]{graphicx}
\usepackage[colorlinks=true,citecolor=blue]{hyperref}
\usepackage{graphicx,textcomp,dcolumn,hyperref,upgreek}

\usepackage{physics}
\usepackage{amssymb}

\def\spin{\boldsymbol{\rho}}
\def\detuning{\eta}
\DeclareMathAlphabet\mathbfcal{OMS}{cmsy}{b}{n}

\usepackage{url}

\usepackage{dsfont}

\usepackage{blindtext}

\usepackage{float}

\DeclareRobustCommand{\neswarrow}{\mathrel{\text{\ooalign{$\swarrow$\cr$\nearrow$}}}}
\DeclareRobustCommand{\nwsearrow}{\mathrel{\text{\ooalign{$\searrow$\cr$\nwarrow$}}}}

\begin{document}

\title{
Magnon kinetic theory of the antiferromagnetic Hanle effect
}
\author{Eric~Kleinherbers}
\email{ekleinherbers@physics.ucla.edu}
\affiliation{Department of Physics and Astronomy and Bhaumik Institute for Theoretical Physics, University of California, Los Angeles, California 90095, USA}
\author{Yaroslav~Tserkovnyak}
\affiliation{Department of Physics and Astronomy and Bhaumik Institute for Theoretical Physics, University of California, Los Angeles, California 90095, USA}
               
\date{\today}

\begin{abstract} 
Motivated by the recently discovered magnonic Hanle effect in an insulating antiferromagnet~\cite{wimmer_2020}, we develop a spin transport theory based on low-energy waves of antiferromagnetic Néel order. These waves have two polarizations, which we describe in analogy to optics using the Stokes vector on the Poincaré sphere. We find that the polarization, which encodes the magnon spin angular momentum, changes periodically with a frequency that is nonlinear in the magnetic field. This explains the observed asymmetry in the Hanle signal as a function of the magnetic field, along with other salient experimental features. By providing an energy-resolved description of the spin injection, our theory combines the kinetic transport of magnons with the coherent dynamics of their polarization in an intuitive way. This opens a general perspective on a coherent control of magnonic spin density in collinear antiferromagnets.
\end{abstract}

\maketitle

Antiferromagnetic materials, characterized by their staggered spin order (Néel order), have become promising candidates in spintronic applications~\cite{baltz_2018,xiong_2022}.
Due to their net zero magnetization and ability to harness spin-orbit effects~\cite{wadley_2016,duttagupta_2020,park_2011,troncoso_2020,das_2022,marti_2023,wang_2023}, they possess many useful properties, such as the lack of stray magnetic fields,  the resilience to external magnetic noise, and ultrafast spin dynamics~\cite{kimel_2004}, while retaining effective magnetoelectric controls.  
\begin{figure}[t]
\includegraphics[width=9cm]{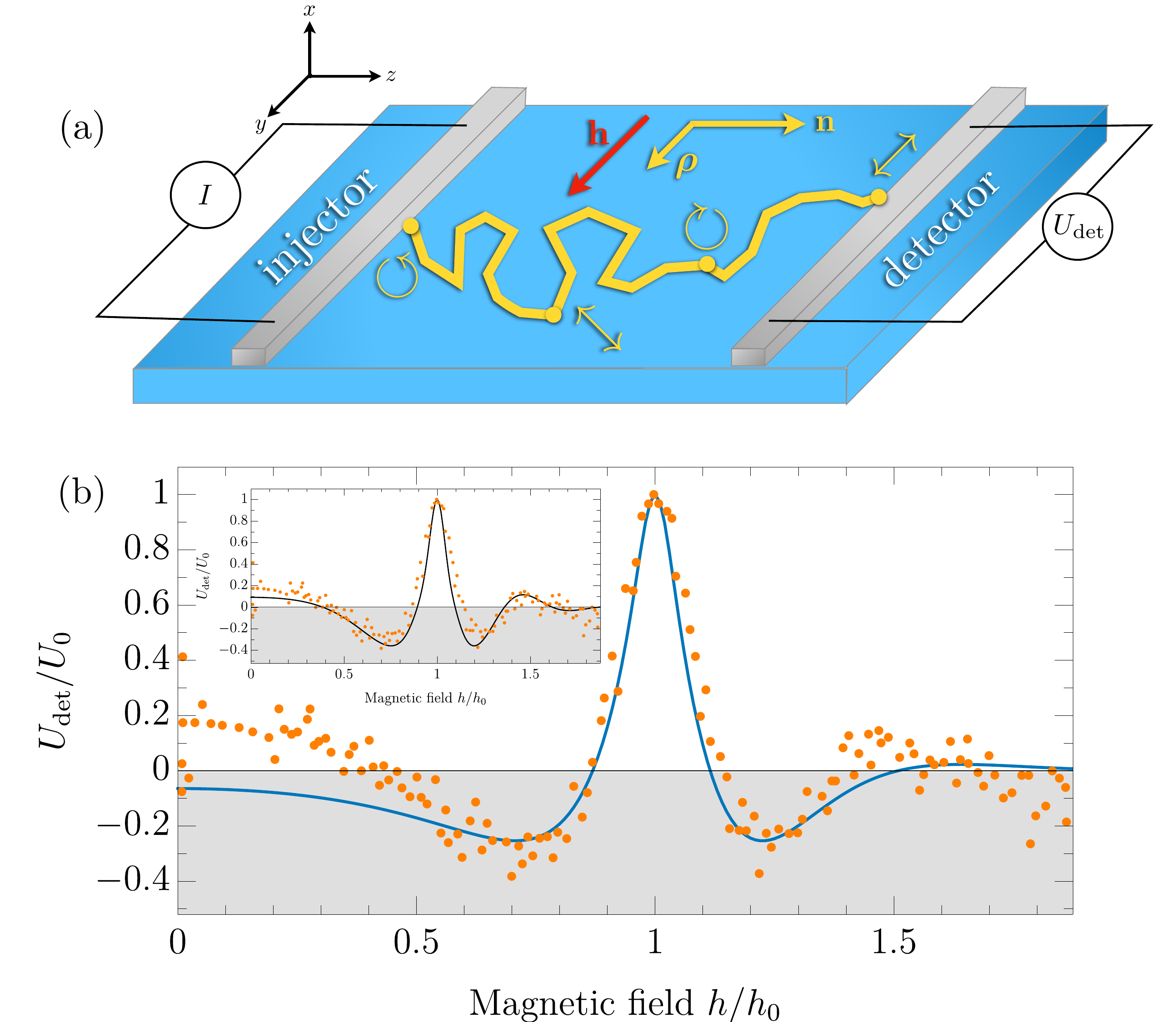}
	\caption{(a) Experimental geometry. Antiferromagnetic hematite (blue) is in contact with two platinum electrodes (gray) separated by length $\ell$ which serve as spin injector and detector, respectively. An injected right-circularly polarized Néel order wave continuously changes polarization (and thus spin) along the trajectory (yellow).  
 (b) Measured spin Hall signal $U_\text{det}$ normalized by the maximum $U_0$ as a function of the in-plane magnetic field $h$ normalized by $h_0$ [see Eq.~\eqref{eq:h0}]. Experimental data (orange dots) were read off from Fig.~2 in  Ref.~\cite{wimmer_2020} and are compared to our theory (line). The parameters for the fit are  $\tau_\text{rel}=80\,h_0^{-1}$, ${\cal D}=0$, $\ell=5\,\sqrt{D \tau_\text{rel}}$, $k_\text{B}T=30\,h_0$, and $\Delta=4\,h_0$. For the inset, we use  the parameters $\tau_\text{rel}=23\,h_0^{-1}$ and $\ell=22\,\sqrt{D \tau_\text{rel}}$.}
	\label{fig:1}
\end{figure}
The staggered character of the spin order leads to \textit{two} kinds of low-energy excitations (magnons) that are distinguished by their polarization~\cite{daniels_2018,nambu_2020}. This polarization, in turn, controls the spin angular momentum of the magnons, which can provide a handle on the spin current in antiferromagnetic devices~\cite{wimmer_2020,han_2020,hortensius_2021,han_2023}. Recently, this has been demonstrated in the form of the \textit{magnonic Hanle effect}~\cite{wimmer_2020}, where the magnon spin is manipulated by an applied magnetic field~\footnote{We remark an important difference to the original Hanle effect in optics~\cite{hanle_1924} describing the changed polarization of resonance fluorescence of a dilute gas due to a magnetic field. There, the rotation of the polarization is \textit{independent} of the time the wave travels through the magnetic field. In this respect, here, the effect is more similar to the Faraday effect.}. 
In this experiment, see Fig.~\ref{fig:1}(a), a spin current is continuously injected into the antiferromagnet (blue) via the spin Hall effect ($I$) and measured at the detector via the inverse spin Hall effect ($U_\text{det}$). The resultant spin Hall voltage $U_\text{det}$ shows a maximum at a finite magnetic field $h_0$, two minima, and an asymmetry with respect to $h_0$,  see Fig.~\ref{fig:1}(b). 
In Ref.~\cite{wimmer_2020}, the experimental findings are interpreted in analogy to the \textit{electronic spin Hanle effect}~\cite{jedema_2002}, i.e., in terms of the diffusive transport of a pseudospin~\cite{wimmer_2020,glueckelhorn_2022,glueckelhorn_2023,kamra_2020} that undergoes a Larmor-like precession in a pseudomagnetic field. While this phenomenological theory does indeed capture some essential features, it lacks a more fundamental, energy-resolved description. Moreover, it does not explain the asymmetry of the experimental signal and crucially depends on the presence of Dzyaloshinskii–Moriya interaction (which we argue not to be the case).

In this paper, we develop a microscopic perspective, which bridges the key aspects of magnon kinetics with coherent precession of the Stokes vector parametrizing elliptical polarization, while accounting for energy dependence of the spin injection, propagation, and detection processes.
To this end, we start with a classical description of the dynamics of the Néel order waves and represent their polarization using  the \textit{Stokes vector} on the \textit{Poincaré sphere} in analogy to optics. Employing an energy-resolved description of the interfacial spin currents~\cite{bender_2015,reitz_2020}, we express the spin Hall signal at the detector as
\begin{align}\label{eq:spinhall}
    U_\text{det}\approx{\cal C} \int \limits_{\Delta}^{k_\text{B}T} \mathrm{d}\varepsilon \int\limits_0^\infty \mathrm{d}\tau g(\ell,\tau) \cos\left[\detuning(\varepsilon) \tau\right].
\end{align}
The result has an intuitive interpretation, where each magnon of energy $\varepsilon$ (the lower bound $\Delta$ is roughly given by the magnon gap and the upper bound by temperature $k_\text{B}T$) travels diffusively from injector ($z=0$) to detector ($z=\ell$) in a time $\tau$ described by the diffusive propagator $g(\ell,\tau)$. The polarization and thus spin of the magnon oscillates as $\cos\left[ \eta(\varepsilon)\tau\right]$, where the frequency $\eta(\varepsilon)$ depends nonlinearly on the magnetic field $h$ and the energy $\varepsilon$. 
In Fig.~\ref{fig:1}(b), we compare the theory (blue line) to the detected spin signal (orange dots). The equation~\eqref{eq:spinhall} is derived below, along with the expression for the prefactor ${\cal C}$. We remark that additional energy-dependent factors in the integrand cancel out in the low-energy description of a collinear antiferromagnet.

\textit{Néel order dynamics.}---Classically, the magnetic order in an antiferromagnet is conveniently described by two fields, namely the Néel order $\vb{n}$ with $\vert \vb{n} \vert=1$ and the spin density $\spin$ with $\vb{n}\cdot\spin=0$. The Lagrangian takes the form~\cite{takei_2014}
\begin{align}
{\cal L}\lbrack \vb{n},\spin\rbrack=& \spin \cdot {\vb{n}}\times\dot{\vb{n}}- \frac{A}{2}\sum_{i=x,y,z}\left(\partial_i \vb{n} \right)^2- \frac{\spin^2}{2\chi} \nonumber \\ &+\spin \cdot \left(\vb{h}+\vb{n}\times{{\mathbfcal{D}}}\right)- \frac{K}{2} \left(\vb{c}\cdot \vb{n}\right)^2,  
\end{align}
where $A$ is the Néel order stiffness and $\chi$ the transverse spin susceptibility. 
In addition, we include Dzyaloshinskii–Moriya interaction with $\mathbfcal{D}={\cal D}\vb{e}_x$  and a  Zeeman term with the externally applied magnetic field  $\vb{h}=h\vb{e_y}$. Also, the system possesses an easy-plane anisotropy $K$ with $\vb{c}=\vb{e}_x$~\cite{wimmer_2020}.
To incorporate Gilbert damping, we use the Rayleigh dissipation function~\cite{hals_2011}
\begin{align}
    {\cal R}[\dot{\vb{n}}]= \frac{\alpha}{2} \dot{ \vb{n}}^2,
\end{align}
with Gilbert damping coefficient $\alpha$.
We study the classical wave dynamics of an injected right-circularly polarized Néel mode (which is the classical counterpart of magnons with spin $+\hbar$). For this purpose, we integrate out the spin density $\spin$ and expand in small perturbations $\vb{n}=\vb{n}_0+ \delta\vb{n}$
with $\vb{n}_0=(0,0,1)$ and $\delta\vb{n}=(\delta n_x, \delta n_y ,-(\delta n_x^2+\delta n_y^2)/2)$ ensuring $\vb{n}^2=1$.
We find the decoupled wave equations of two linearly polarized Néel modes~\cite{SM}
\begin{align}
\left(\partial_t^2-c^2 \grad^2 + \frac{1}{\tau_\text{rel}} \partial_t+ \Delta_{x/y}^2\right) \delta n_{x,y}=0,
\end{align}
where we define the magnon speed $c=\sqrt{A/\chi}$, the relaxation time $\tau_\text{rel}=\chi/\alpha$, and the frequency gaps $\Delta_x^2=K/\chi + {\mathcal{D}}^2+ h{\mathcal{D}}$ and $\Delta_y^2=h^2+h{\mathcal{D}}$.
The resulting eigenmodes have gapped dispersions $\omega_{x/y}= \sqrt{\Delta_{x/y}^2+(ck)^2}+{\cal O}(\tau_\text{rel}^{-2})$ which give rise to the average frequency and the detuning
\begin{align}
\bar{\omega}(k)&= \frac{\omega_x(k)+\omega_y(k)}{2},\\
\detuning(k) &= {\omega_x(k)-\omega_y(k)},\label{eq:detuning}
\end{align}
respectively. 
The plane-wave solutions (up to taking the real part) are then given by 
\begin{align}\label{eq:jones}
\delta \vb{n}(\vb{r},t)=C \vb{p}(t)e^{i\left[\vb{k}\cdot\vb{r}-\bar{\omega}(k) t\right]}e^{-{t}/{2\tau_\text{rel}}},
\end{align}
with small amplitude $C\ll1$.  The polarization is described by the (complex) Jones vector 
\begin{align}
\vb{p}(t)=\frac{1}{\sqrt{2}}\left[{e^{-i{\detuning(k)} t /2}}\vb{e}_{x}+ {ie^{i{\detuning(k)} t/2}} \vb{e}_{y}\right]\label{eq:polarization_jones}
\end{align}
which, initially, is right circular, $\vb{p}(0)=\vb{p}_\circlearrowleft=(\vb{e}_{x} +i \vb{e}_{y})/\sqrt{2}$ and thus a superposition of linear polarized eigenmodes. 
A convenient tool to visualize the polarization is given by the (real) 3d Stokes vector $\vb{S}=(S_1,S_2,S_3)$ defined on the Poincaré sphere~\cite{malykin_1997}. It can be derived from $\vb{p}(t)$ via
\begin{align}
\vb{S}(k,t)={ R}_1\left(\frac{\pi}{2}\right)\sum_{i,j}{{p}}_i^*(t)\boldsymbol{\sigma}_{ij}{{p}}_j(t),\label{eq:def_stokes}
\end{align} 
using the Pauli matrices $\boldsymbol{\sigma}=(\sigma_1,\sigma_2,\sigma_3)$. To ensure that the north (south) pole of the Poincaré sphere corresponds to a right-circular (left-circular) polarization $\vb{p}_\circlearrowleft$ ($\vb{p}_\circlearrowright$), we perform a rotation by $\pi/2$ about the $1$-axis using ${R}_1({\pi}/{2})$. We emphasize that the Stokes vector is, in principle, not related to real space. Only its projection on the 3-axis, $\vb{e}_3\cdot \vb{S}(k,t)$, is proportional to the spin (projected on the Néel order $\vb{n_0}$) that is carried by the wave~\cite{SM}. Thus, linear polarized waves found on the equator of the Poincaré sphere have zero spin~\cite{liensberger_2019}. 
\begin{figure}[t]	\includegraphics[width=9cm]{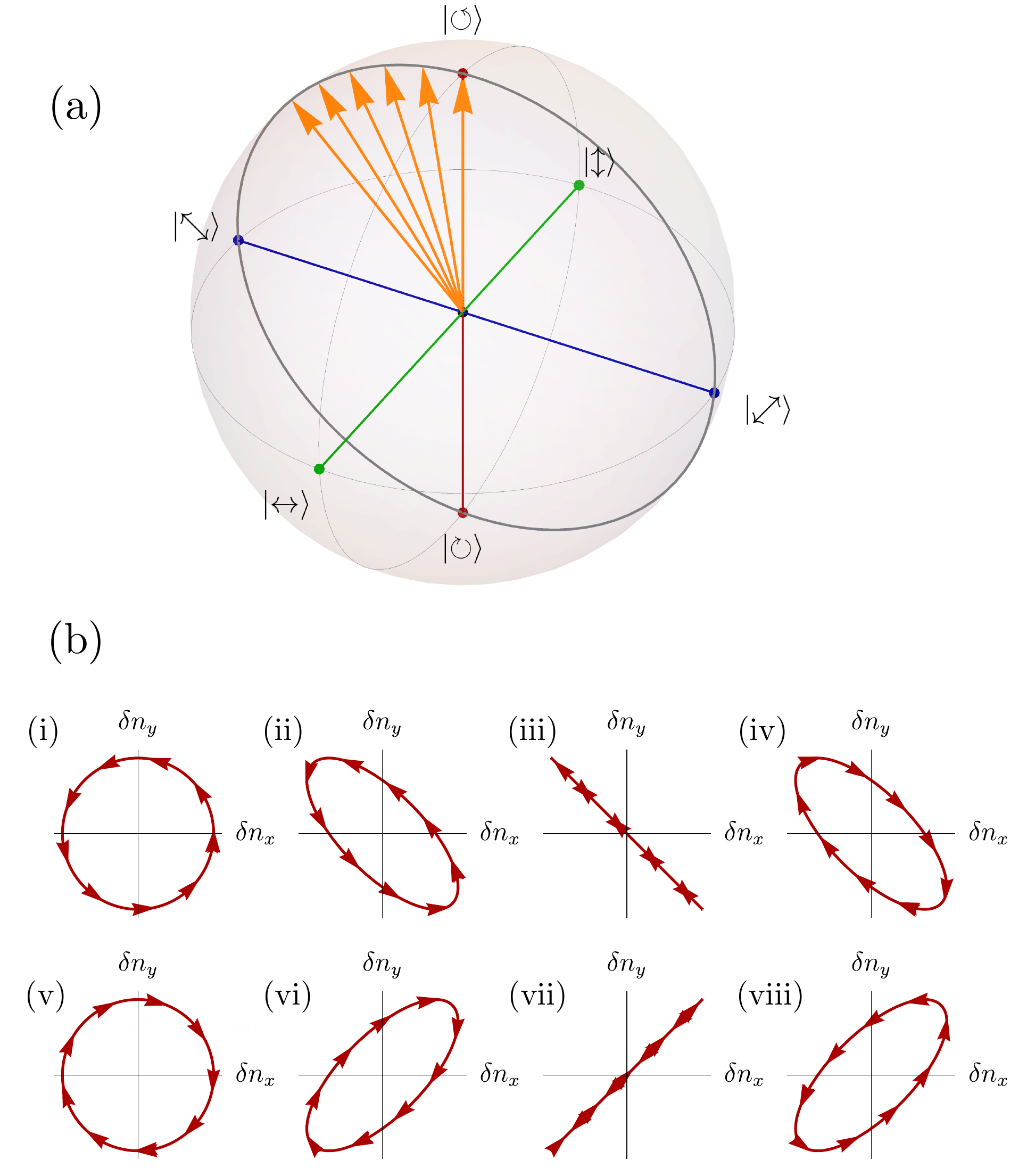}
	\caption{(a) Polarization of Néel order mode described by 3d Stokes vector (orange arrow) which changes on the Poincaré sphere by rotation around the 2-axis (green), where $\ket{\leftrightarrow}$ and $\ket{\updownarrow}$ are the stationary  polarizations of the eigenmodes. (b) Néel mode polarization changing from (i) right circular $\ket{\circlearrowleft}$ to (iii) $-\pi/4$ linear  $\ket{{\nwsearrow}}$ to (v) left circular $\ket{\circlearrowright}$ to (vii) $+\pi/4$ linear $\ket{\neswarrow}$ and back to (i) right circular $\ket{\circlearrowleft}$. In between [(ii),(iv),(vi),(viii)], the polarization is elliptical.}
	\label{fig:polar}
\end{figure}
From Eq.~\eqref{eq:polarization_jones} and Eq.~\eqref{eq:def_stokes}, we find that the dynamics of the Stokes vector are described by 
\begin{align}\label{eq:pol}
\dot{\vb{S}}(k,t)= -\boldsymbol{\detuning}(k)\times \vb{S} ,
\end{align}
where $\boldsymbol{\detuning}(k)=(0,\detuning(k),0)$. Thus, the detuning $\boldsymbol{\detuning}(k)$  leads to a precession of the Stokes vector around the $2$-axis on the Poincaré sphere which corresponds to a continuously changing polarization of the wave, see Fig.~\ref{fig:polar}(a)-(b). Specifically, the polarization is changing from (i) right circular to (iii) $-\pi/4$ linear to (v) left circular to (vii) $+\pi/4$ linear and back to right circular (i).

From our classical considerations of Néel dynamics, we can draw already two of the main conclusions about the magnonic Hanle effect. First, for a particular nonzero magnetic field 
\begin{align}\label{eq:h0}
h_0=\sqrt{K/\chi+{\cal D}^2},
\end{align}
we get $\Delta_x=\Delta_y$, i.e., a degeneracy of the two linearly-polarized eigenmodes $\delta n_{x,y}$~\footnote{By calculating the dispersion for a simple cubic lattice using the Holstein-Primakoff representation of spins, we find that this is actually true for the full Brillouin zone.}. In this limit, there is zero precession, $\boldsymbol{\detuning}(k)=0$, and circularly polarized modes become proper eigenmodes as well. This, in turn, means that the spin angular momentum of an injected Néel order wave does not oscillate in time resulting in a maximum signal $U_\text{det}$ at the detector, see Fig.~\ref{fig:1}(b). Second, the precession frequency is, in general, nonlinear in the magnetic field
\begin{align}
    \detuning(k)&=\sqrt{h_0^2+h{\cal D} +(ck)^2}-\sqrt{h^2+h{\cal D} +(ck)^2} \label{eq:dephase_freq}\\
    &= \frac{h_0^2-h^2}{2ck} +{\cal O}\left(\frac{1}{k^{3}}\right),
\end{align}
where in the second line we performed an expansion in $k^{-1}$.
Due to this nonlinearity in $h$, the local minima of the signal $U_\text{det}$ are found nonsymmetrically around $h_0$ at positions $h_0\pm \delta h_\pm $ with $\delta h_- \neq \delta h_+$. Using the condition $\detuning(k) \tau=\mp \pi$ that the polarization (spin) has flipped from right-circularly ($+\hbar$) to left-circularly ($-\hbar$) polarized modes, where $\tau$ is a characteristic time for a magnon to travel from injector to detector, we find that $\delta h_- \ge \delta h_+$. 
This explains why  the signal $U_\text{det}$ is stretched for $h<h_0$ and compressed for $h>h_0$, see Fig.~\ref{fig:1}(b). Only for ${\cal D}=k=0$, we retrieve Larmor-like physics with $\detuning(0)=h_0-h$ and the signal becomes symmetric, $\delta h_-= \delta h_+$.  Moreover, we find that the Dzyaloshinskii–Moriya interaction $\cal D$ is dispensable for the effect in contrast to what is found in Ref.~\cite{wimmer_2020}. 

\begin{figure}[t]	\includegraphics[width=9cm]{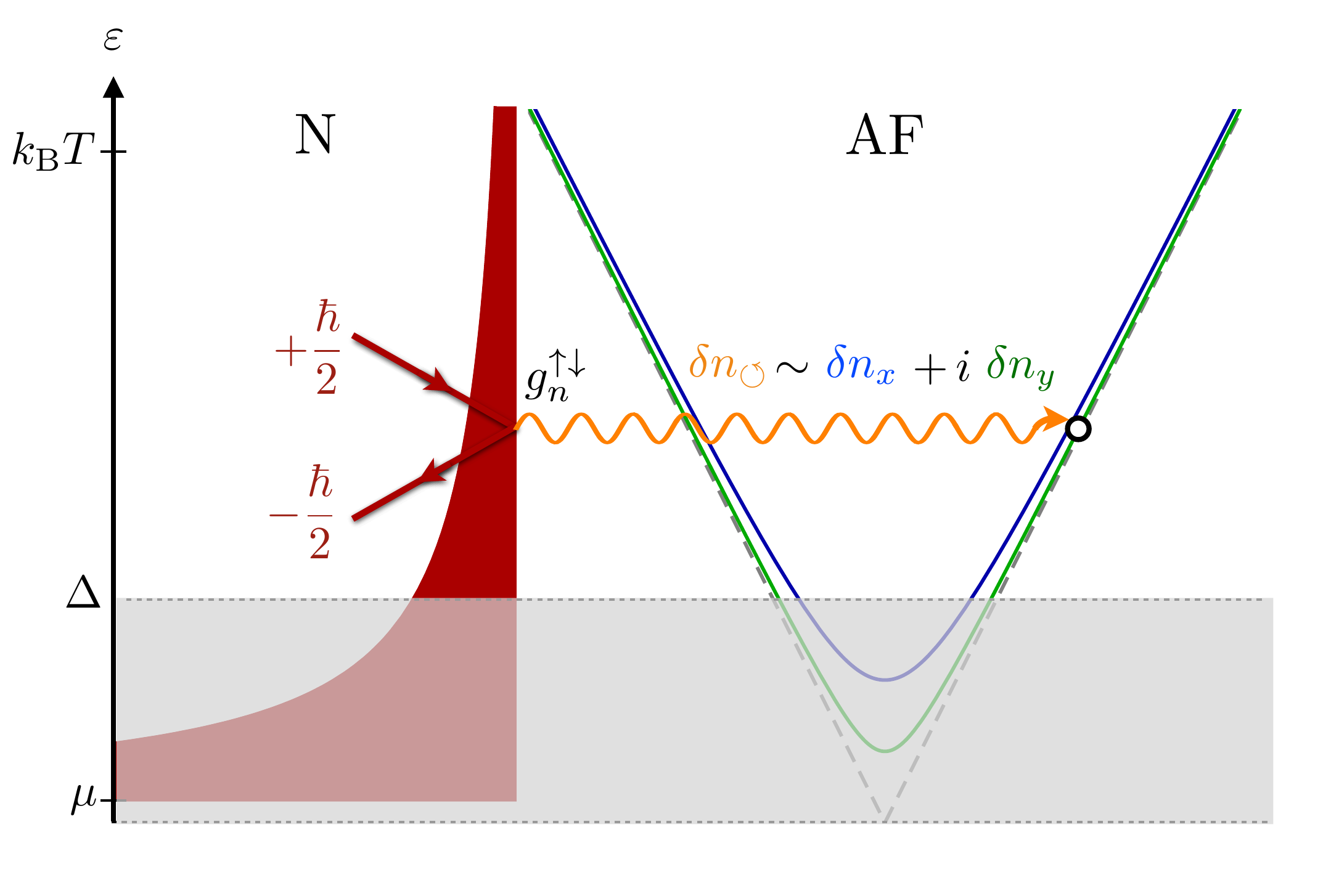}
	\caption{Spin injection at the interface between normal metal (N) and antiferromagnet (AF). Spin-flip excitations in N are effectively described by a Bose-Einstein distribution (red) with spin accumulation $\mu$ and temperature $T$. At the interface, electrons flip their spin from $+\hbar/2$ to $-\hbar/2$ (red arrows) creating a right-circularly polarized Néel mode (orange arrow), i.e., a superposition of eigenmodes (green and blue). Due to a low velocity and high detuning, excitations close to the band gap (gray area) can be neglected by approximating $\varepsilon=\hbar ck$ (gray dashed line) and limiting the integration from $\Delta$ to $k_\text{B}T$.}
	\label{fig:injection}
\end{figure}

\textit{Kinetic theory.}---To quantify the time $\tau$  it takes for magnons to travel from injector to detector, we assume for each energy $\varepsilon$ a 2d diffusive transport regime~\cite{manchon_2017} with $\vb{r}=(y,z)$ due to the effective 2d geometry of the setup
\begin{align}\label{eq:diff}
\partial_t m(\varepsilon,\vb{r},t)= D \grad^2  m -\frac{m}{\tau_{\text{rel}}}, 
\end{align}
with magnon density $m(\varepsilon,\vb{r},t)$ per volume and energy, diffusion constant $D$, and spin relaxation time $\tau_\text{rel}$. Here, we assume besides Gilbert damping also elastic, spin-conserving scattering events with a constant mean free path $\ell_\text{mfp}$, leading to an energy-independent diffusion coefficient $D\sim c \,\ell_\text{mfp}$.
We remark that the kinetic theory is readily adaptable to other transport regimes.  
Now, in order to keep track of the polarization dynamics, we combine Eq.~\eqref{eq:pol} and Eq.~\eqref{eq:diff} to obtain 
\begin{align}\label{eq:mastereq}
\partial_t{\vb{q}}(\varepsilon,\vb{r},t)
=D \grad^2 \vb{q} -\frac{\vb{q}}{\tau_{\text{rel}}}-\boldsymbol{\detuning}(\varepsilon)\times \vb{q} +{\vb{Q}(\varepsilon,{\vb{r}}}),
\end{align} 
where $\vb{q}(\varepsilon,\vb{r},t)$
is the coarse-grained product of the magnon density $m(\varepsilon,\vb{r},t)$ and the associated Stokes vector $\vb{S}(\varepsilon,\vb{r},t)$ describing the polarization. Here, we approximate the dispersion as linear $\varepsilon\approx \hbar c k$ and write $\boldsymbol{\detuning}(\varepsilon)\equiv\boldsymbol{\detuning}(k)$. For consistency, we introduce a lower cutoff $\Delta<\varepsilon$ with $\Delta\sim \Delta_{x,y}$, anticipating that modes near the gap are of minor importance in spin transport due to their low velocity and high detuning, see Fig.~\ref{fig:injection}.  Finally, we added a source term 
\begin{align}
\vb{Q}(\varepsilon,{\vb{r}})=j_{s}^\text{inj}(\varepsilon) \frac{W}{d} \delta(z) \Theta\left(\frac{L}{2}-\vert y \vert\right) \vb{e}_3
\end{align}
which continuously injects a spin current $I^\text{inj}_{s}=\int\mathrm{d}\varepsilon \,j^\text{inj}_{s}(\varepsilon) W L$ at $z=0$ of right-circularly polarized Néel order waves at the interface. Here, $W$ and $L$ are the width and length of the injector and $d$  is the thickness of the hematite film. As a next step, we estimate the energy dependence of the injected spin current $j_{s}^\text{inj}(\varepsilon)$.

\begin{figure}[t]	\includegraphics[width=9cm]{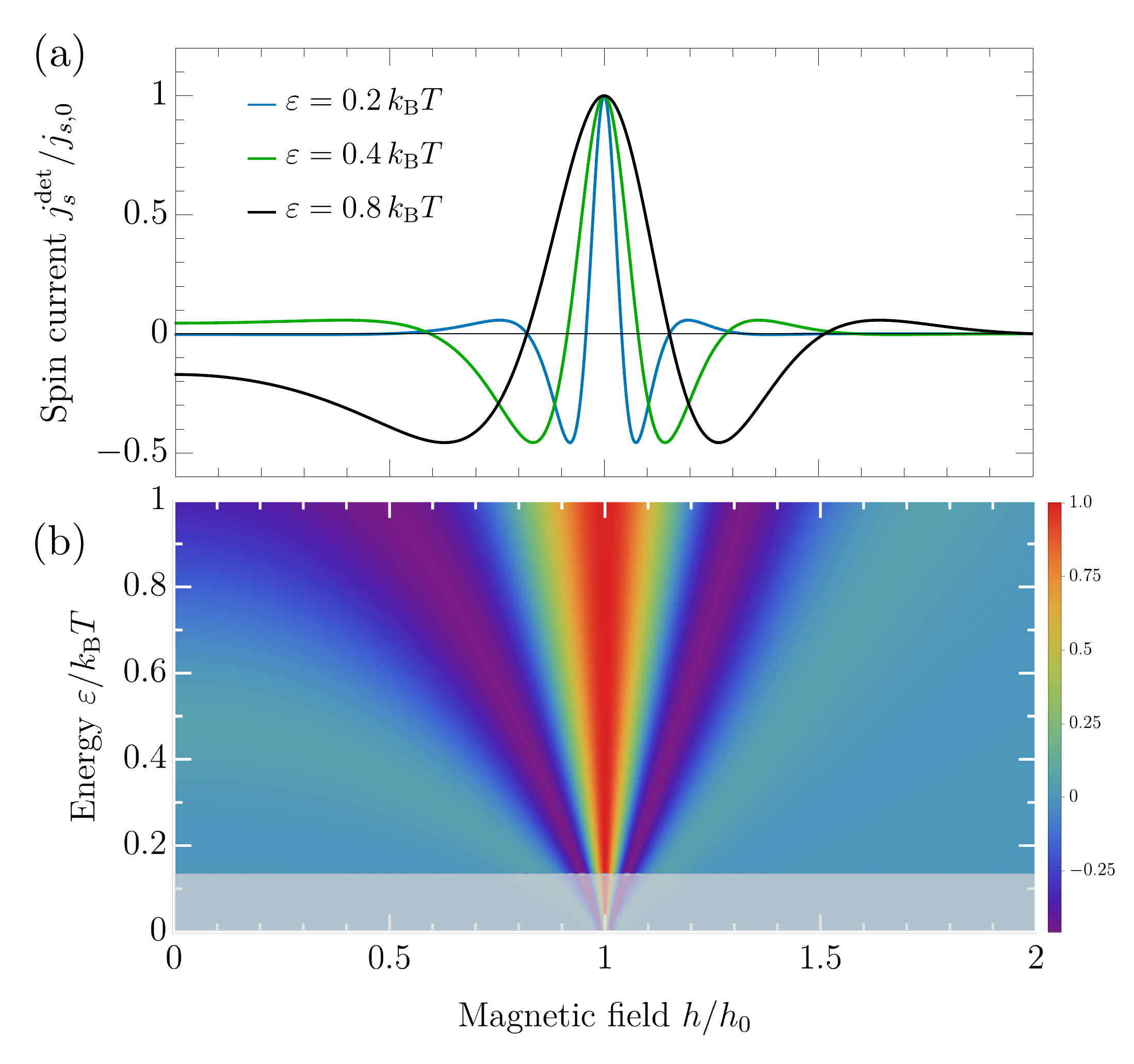}
	\caption{(a) Spin current density $j_s^\text{det}(\varepsilon)$ per energy injected into the detector as a function of magnetic field $h$ for $\varepsilon=0.2\,k_\text{B}T$ (blue), $\varepsilon=0.4\,k_\text{B}T$ (green) and $\varepsilon=0.8\,k_\text{B}T$ (black). $j_s^\text{det}(\varepsilon)$ is normalized by the maximum value $j_{s,0}$. In (b), we show $j_s^\text{det}/j_{s,0}$ as a function of both $h$ and $\varepsilon$. Since we use an approximated linear dispersion $\varepsilon\approx c k$, the excitations close to the band gap (gray shaded area) are not adequately captured. The remaining parameters are the same as in Fig.~\ref{fig:1}.}
\label{fig:energyresolved}
\end{figure}

\textit{Interfacial spin current.}---At the interface, itinerant electrons in the normal metal can inelastically flip their spin $\hbar/2$ to create a magnon (right-circularly polarized Néel mode) in the antiferromagnet and vice versa, see Fig.~\ref{fig:injection}.
To model this interfacial spin current, we use the formula~\cite{SM}
\begin{align}
    j_s(\varepsilon)=\frac{\hbar^2 g_n^{\uparrow \downarrow}}{4\pi \chi} {\cal N}(\varepsilon)\left[n_\text{BE}(\varepsilon-\mu)-n(\varepsilon)\right],
\end{align}
where $n_\text{BE}(\varepsilon{-}\mu)=(e^{ (\varepsilon{-}\mu)/k_\text{B}T}-1)^{-1}$ describes the Bose-Einstein distributed spin-flip excitations in the metal with spin accumulation $\mu$, and $n(\varepsilon)$ describes the magnon occupation in the antiferromagnet. Here,
$g^{\uparrow\downarrow}_n$ is the interfacial spin-mixing conductance per area which we assume to be equal at injector and detector. ${\cal N}(\varepsilon)=\frac{\varepsilon^2}{2\pi^2c^3\hbar^3}$ is the density of states per energy and volume.  
We emphasize the  difference to a normal metal/ferromagnet interface~\cite{bender_2015}, where we would obtain an additional factor of $\varepsilon$. In the following, we assume that the interfaces act as bottlenecks of spin transfer so that backflow is negligible. 
Thus, at the injector, we can assume
$n(\varepsilon)\approx n_\text{BE}(\varepsilon)$ leading to the inflowing spin current
\begin{align}
j^\text{inj}_s(\varepsilon)\approx \frac{g^{\uparrow\downarrow}_n}{(2\pi c)^3 \hbar \chi} \mu \,k_\text{B}T,
\end{align}
where we expanded in the spin accumulation $\mu$ and assumed the Rayleigh-Jeans limit ($\varepsilon \ll k_\text{B}T$). Remarkably, the current becomes completely energy independent.
Similarly, at the detector, where $\mu=0$, the excess magnons are now on the antiferromagnetic side and given by the diffusive spin density $\rho_z(\varepsilon,\vb{r})$ at $\vb{r}=(y,\ell)$ with $-L/2<y<L/2$. We obtain the outflowing spin current
\begin{align}
j^\text{det}_s(\varepsilon,\vb{r})=  \frac{\hbar g^{\uparrow\downarrow}_n}{4\pi\chi}  {\rho_z(\varepsilon,\vb{r})},
\end{align}
where $\rho_z(\varepsilon,\vb{r})=\hbar \vb{e}_3\cdot \vb{q}(\varepsilon,\vb{r})$ is given by the stationary solution of Eq.~\eqref{eq:mastereq}~\cite{SM}.
In Fig.~\ref{fig:energyresolved}(a), we show the normalized spin current density $j_s^\text{det}(\varepsilon)$ for various energies $\varepsilon$ as a function of the magnetic field $h$.  We find that for all energies the central peak is found at $h=h_0$. However, its width  is roughly given by
\begin{align}
    \Delta h  \approx \frac{2\pi}{\hbar}\sqrt{\frac{ D}{l \tau_\text{rel}}}\frac{\varepsilon}{h_0},
\end{align} 
and therefore grows linearly with energy $\varepsilon$, see Fig.~\ref{fig:energyresolved}(b). Additional, secondary peaks occur whenever the Stokes vector $\vb{S}$ of the magnons rotates on average by an angle $2n\pi$ (maxima) or $(2n+1)\pi$ (minima) around the Poincaré sphere on their way from injector to detector. Since the detuning $\eta(\varepsilon)$ is energy dependent, the secondary peak positions change as a function of $\varepsilon$.  Also, the asymmetry in the magnetic field $h$ is clearly visible in Fig.~\ref{fig:energyresolved}(b).

Finally, in the detector, 
this spin current $j^\text{det}_s$ in turn induces a charge current $j^\text{det}_c$ via the inverse spin Hall effect, $j^\text{det}_c=\theta j^\text{det}_s$,  which gives rise to the detected voltage 
\begin{align}
    U_\text{det}& =\frac{\theta}{\sigma}\int\limits_{-L/2}^{L/2} \mathrm{d}y \int \mathrm{d}\varepsilon \,j^\text{det}_s(\varepsilon,y,\ell) \nonumber\\ \label{eq:udet}
    &\approx \underbrace{\frac{\theta}{\sigma} \frac{ \hbar^2 g^{\uparrow\downarrow}_n}{4\pi \chi} \frac{WL}{d} j_s^\text{inj}}_{{\cal C}}\int \limits_{\Delta}^{k_\text{B}T} \mathrm{d}\varepsilon \int\limits_0^\infty \mathrm{d}\tau g(\ell,\tau) \cos\left[\detuning(\varepsilon) \tau\right],
\end{align}
where $\theta$ is proportional to the spin Hall angle~\cite{keller_2018} and $\sigma$ is the electrical conductivity. In the second line, we use that $L\gg \ell$ and introduce the propagator $g(\ell,\tau)=e^{-{\ell^2}/{4D\tau}} e^{-\tau/\tau_\text{rel}}/{\sqrt{4\pi D\tau}}
$ of an effective 1d diffusion equation with relaxation~\cite{SM}. Also, we roughly limit the integration to $\Delta<\varepsilon<k_\text{B}T$, see Fig.~\ref{fig:injection}. The upper limit allows us to use the Rayleigh-Jeans limit, while the lower limit $\Delta\gtrsim \Delta_{x,y}$ ignores states near the band gap due to their slow velocity and large detuning.
The final equation~\eqref{eq:udet} has a very intuitive interpretation. Each magnon with energy $\varepsilon$ that is injected at the source $z=0$ propagates diffusively to $z=\ell$ in a time $\tau$ during which the spin oscillates with $\eta(\varepsilon)$ and decays with $\tau_\text{rel}$. Interestingly,  each energy slice $\mathrm{d}\varepsilon$ contributes equally. By comparing to the experimental data extracted from Ref.~\cite{wimmer_2020}, we find a good agreement, see Fig.~\ref{fig:1}(b). 
By assuming larger separations $\ell\gg\sqrt{D \tau_\text{rel}}$ [see the inset of Fig.~\ref{fig:1}(b)], we can even reproduce the overshooting for small and large magnetic fields, albeit at the expense of a smaller signal.

\textit{Discussion and outlook.}---
Exploiting the coherent control of the polarization of magnons in antiferromagnets is a promising route in future spintronics applications~\cite{wimmer_2020,han_2020,hortensius_2021,han_2023}. 
Starting from the classical Néel order dynamics, we derived an intuitive kinetic equation keeping track of both the magnon density and polarization. 
While the dynamics of the magnon density can be described by a generic kinetic theory, the magnon polarization dynamics is governed by energetics and is described in terms of the Stokes vector on the Poincaré sphere --- in analogy to optics. In addition, we explicitly model the energy-resolved interfacial spin currents, allowing us to describe both spin injection and detection. 
Applying our theory to the magnetic Hanle effect, we found that the injected right-circularly polarized modes show a precession of the Stokes vector with a frequency that is nonlinear in the magnetic field and only weakly dependent on the Dzyaloshinskii–Moriya interaction.
For simplicity, we assumed in our kinetic theory only elastic scattering events leading to an energy-conserving diffusion equation. 
We find that the resulting inverse spin-Hall voltage measured in Ref.~\cite{wimmer_2020} (data extracted from Fig.~2 of publication) is well reproduced by our model, including the asymmetry in the magnetic field.

\begin{acknowledgments}
This work was supported by the U.S.
Department of Energy, Office of Basic Energy Sciences
under Grant No. DE-SC0012190.
\end{acknowledgments}

\bibliography{References}

\end{document}


\title{
Supplementary for
``Magnon kinetic theory of the antiferromagnetic Hanle effect"
}
\author{Eric~Kleinherbers}
\email{ekleinherbers@physics.ucla.edu}
\affiliation{Department of Physics and Astronomy and Bhaumik Institute for Theoretical Physics, University of California, Los Angeles, California 90095, USA}

\author{Yaroslav~Tserkovnyak}
\affiliation{Department of Physics and Astronomy and Bhaumik Institute for Theoretical Physics, University of California, Los Angeles, California 90095, USA}
               
\date{\today}

\maketitle


\section{System}
Classically, the magnetic order in an antiferromagnet is conveniently described by two fields, namely the Néel order $\vb{n}$ with $\vert \vb{n} \vert=1$ and the spin density $\spin$ with $\vb{n}\cdot\spin=0$. Here, the Lagrangian is given by~\cite{takei_2014}
\begin{align}
{\cal L}\lbrack \vb{n},\spin\rbrack=& \spin \cdot {\vb{n}}\times\dot{\vb{n}}- \frac{A}{2}\sum_{i=x,y,z}\left(\partial_i \vb{n} \right)^2- \frac{\spin^2}{2\chi} \nonumber \\ &+\spin \cdot \left(\vb{h}+\vb{n}\times{{\mathbfcal{D}}}\right)- \frac{K}{2} \left(\vb{c}\cdot \vb{n}\right)^2,  
\end{align}
where $A$ is the Néel order stiffness and $\chi$ the transverse spin susceptibility. 
In addition, we include Dzyaloshinskii–Moriya interaction with $\mathbfcal{D}={\cal D}\vb{e}_x$  and a  Zeeman term with $\vb{h}= \gamma \vb{H}=h\vb{e_y}$ given by the magnetic field $\vb{H}$ and the gyromagnetic factor $\gamma$. Finally, the system possesses an easy-plane anisotropy $K$ with $\vb{c}=\vb{e}_x$~\cite{wimmer_2020}.
To incorporate Gilbert damping, we use the Rayleigh dissipation function~\cite{hals_2011}
\begin{align}
    {\cal R}[\dot{\vb{n}}]= \frac{\alpha}{2} \dot{ \vb{n}}^2,
\end{align}
with Gilbert damping coefficient $\alpha$.
We eliminate the spin density by formally inserting the equilibrium value which can be obtained via
\begin{align}
    \delta_{\spin} {\cal L}=0 \quad \Rightarrow \quad \spin=\chi \left(\vb{n}\times\vb{\dot{n}} + \vb{h}+ \vb{n}\times \mathbfcal{D} \right).\label{eq:spindens_def}
\end{align}
Then, we arrive at an effective Lagrangian for the Néel order
\begin{align}
{\cal L}\lbrack \vb{n}\rbrack=&\frac{\chi}{2} \left[\dot{\vb{n}}+\vb{n}\times\left(\vb{h}+\vb{n}\times{{\mathbfcal{D}}}\right) \right]^2 \nonumber \\ &- \frac{A}{2}\sum_{i=x,y,z}\left(\partial_i \vb{n} \right)^2 - \frac{K}{2} \left(\vb{c}\cdot \vb{n}\right)^2.   
\end{align}
\section{Néel order wave dynamics}
At the injector, a right circular polarized Néel mode (which is the classical counterpart of a magnon with spin $+\hbar$) is excited. To derive its time evolution in the antiferromagnet, we study small perturbations $\vb{n}=\vb{n}_0+ \delta\vb{n}$
with $\vb{n}_0=(0,0,1)$ and $\delta\vb{n}=(\delta n_x, \delta n_y ,-(\delta n_x^2+\delta n_y^2)/2)$ ensuring $\vb{n}^2=1$. Then, we arrive (up to a total time derivative) at
\begin{align}
{\cal L}\lbrack \delta\vb{n}\rbrack=&\frac{\chi}{2} \left(\dot{\delta n}_x^2+\dot{\delta n}_y^2\right) -\frac{A}{2} \sum_{i=x,y}\left(\partial_i \delta\vb{n}\right)^2 \nonumber \\ &- \frac{\chi}{2}\left(\frac{K}{\chi}+{\mathcal{D}}^2+ h{\mathcal{D}}\right)\delta n_x^2 - \frac{\chi}{2}\left( h^2+h{\mathcal{D}}\right)\delta n_y^2
\end{align}
and 
\begin{align}
 {\cal R}\lbrack \delta\vb{\dot{n}}\rbrack=&\frac{\alpha}{2} \left(\dot{\delta n}_x^2+\dot{\delta n}_y^2\right).
\end{align}
Using the Euler-Lagrange equation
\begin{align}
    \dv{t} \frac{\delta {\cal L}}{\delta \vb{\dot{n}}}=\frac{\delta {\cal L}}{\delta \vb{n}} - \frac{\delta {\cal R}}{\delta \vb{\dot{n}}}, 
\end{align}
which are modified to include dissipation described by ${\cal R}$, we derive the decoupled wave equation of linearly polarized Néel modes
\begin{align}
\left(\partial_t^2-c^2 \grad^2 + \frac{1}{\tau_\text{rel}} \partial_t+ \Delta_{x/y}^2\right) \delta n_{x,y}=0,
\end{align}
where we defined the magnon speed $c=\sqrt{A/\chi}$, the relaxation time $\tau_\text{rel}=\chi/\alpha$, and the frequency gaps $\Delta_x^2=K/\chi + {\mathcal{D}}^2+ h{\mathcal{D}}$ and $\Delta_y^2=h^2+h{\mathcal{D}}$. 
The plane-wave solutions (up to taking the real part) are given by 
\begin{align}
\delta \vb{n}(\vb{r},t)=\left(C_x e^{i\left(\vb{k}\cdot\vb{r}-\omega_x t\right)}\vb{e}_{x}+C_y e^{i\left(\vb{k}\cdot\vb{r}-\omega_y t\right)}\vb{e}_{y} \right)e^{-t/2\tau_\text{rel}},
\end{align}
with small amplitudes $C_{x,y}\ll1$ and gapped dispersions
\begin{align}
    \omega_x&=\sqrt{\Delta_x^2+(ck)^2-\tau_\text{rel}^2}, \\
    \omega_y&=\sqrt{\Delta_y^2+(ck)^2-\tau_\text{rel}^2}, 
\end{align}
with $\vert\vb{k}\vert=k$.
Thus, a right circular polarized  plane wave (which is the classical counterpart of a magnon with spin $+\hbar$) injected at $\vb{r}=t=0$ of the form $\delta \vb{n}(0,0)=C(\vb{e}_{x} +i \vb{e}_{y})/\sqrt{2}$ will change over time according to
\begin{align}
\delta \vb{n}(\vb{r},t)=C \underbrace{\left(\frac{e^{-i\frac{\eta(k)}{2} t}}{\sqrt{2}}\vb{e}_{x}+ \frac{ie^{i\frac{\eta(k)}{2} t}}{\sqrt{2}} \vb{e}_{y} \right)}_{:=\vb{p}(t)}e^{i\left(\vb{k}\cdot\vb{r}-\bar{\omega}(k) t\right)}e^{-t/2\tau_\text{rel}},\label{eq:neelwave}
\end{align}
where we defined the average frequency and the detuning
\begin{align}
\bar{\omega}(k)&= \frac{\omega_x(k)+\omega_y(k)}{2},\\
\eta(k) &= {\omega_x(k)-\omega_y(k)},
\end{align}
respectively. Moreover, we introduced the 2d complex Jones vector $\vb{p}$ (fulfilling $\vb{n}_0\cdot\vb{p}=0$ and $\vb{p}^*\cdot \vb{p}=1$) describing the polarization of the wave. A convenient tool to visualize the polarization is given by the 3d Stokes vector $\vb{S}=(S_1,S_2,S_3)$ defined on the Poincaré sphere~\cite{malykin_1997}. It can be calculated via 
\begin{align}
\vb{S}(k,t)={\cal R}_1\left(\frac{\pi}{2}\right)\sum_{i,j}{{p}}_i^*(t) \boldsymbol{\sigma}_{ij}{{p}}_j(t)
\end{align} 
using the Pauli matrices $\boldsymbol{\sigma}=(\sigma_1,\sigma_2,\sigma_3)$. To ensure that the north (south) pole of the Poincaré sphere corresponds to a right (left) circular polarization, we perform a rotation by $\pi/2$ about the $1$-axis using ${\cal R}_1(\frac{\pi}{2})$. Note that the Stokes vector is, in principle, unrelated to real space. Only its projection on the 3-axis, $\hbar S_3=\hbar \vb{e}_3\cdot \vb{S}$, measures the spin carried by the wave. 
This can be seen by evaluating the spin density $\rho_z$ from Eq.~\eqref{eq:spindens_def} for $\omega\gg h,{\cal D}$ by inserting $\vb{n}=\Re\left(\vb{n_0}+\delta\vb{n}\right)$, i.e. the real part of the Néel order wave from Eq.~\eqref{eq:neelwave}. We find
\begin{align}
    \rho_z=\vb{e}_z\cdot \spin\approx \chi \vb{e}_z\cdot \vb{n}\times\vb{\dot{n}}\approx -\frac{i\bar{\omega}\chi}{2}\vb{e}_z\cdot \delta\vb{n}^*\times\delta\vb{{n}} =\frac{-i\bar{\omega}\chi }{2}C^2 e^{-t/\tau_\text{rel}}\left(p^*_x {p}_y-p^*_y  p_x\right),
\end{align}
where we assumed a constant polarization $\vb{p}$.
At the same time, we have
\begin{align}
    S_3=\vb{e}_3\cdot \vb{S}=\vb{e}_3\cdot{\cal R}_1\left(\frac{\pi}{2}\right)\sum_{i,j}{{p}}_i^* \boldsymbol{\sigma}_{ij}{{p}}_j= \sum_{i,j}{{p}}_i^* (\sigma_y)_{ij}{{p}}_j=-i\left(p^*_x {p}_y-p^*_y  p_x\right),
\end{align}
resulting in the relation between the spin density and the polarization given by
\begin{align}
    \rho_z=\frac{\bar{\omega}\chi}{2} C^2 e^{-t/\tau_\text{rel}} S_3.\label{eq:spinvspolarization}
\end{align}
Thus, linear polarized waves  with $S_3=0$ have zero spin, $\rho_z=0$~\cite{liensberger_2019}. 

Employing the Stokes vector, the detuning $\detuning(k)$ has now a simple geometric interpretation: it simply rotates the Stokes vector via $\vb{S}(k,t)={\cal R}_2[-\eta(k)  t]\vb{S}(k,0)$ as a function of time on a great circle on the Poincaré sphere which corresponds to  a continuously changing polarization of the wave, see Fig.~2. The  dynamics are described by
\begin{align}\label{eq:pol}
\dot{\vb{S}}(k,t)=-\boldsymbol{\eta}(k)\times \vb{S},
\end{align}
where $\boldsymbol{\eta}=(0,\eta,0)$. Since the polarization and thus the carried spin of the wave depends on time $t$, $\rho_z\sim\cos(\eta t)$, it is crucial to know how long it takes a magnon to travel from the injector to the detector.

\section{Magnon kinetic equation}

To quantify the time $\tau$  it takes for magnons to travel from injector to detector, we assume for each energy $\varepsilon$ a 2d diffusive transport regime with $\vb{r}=(y,z)$ due to the effective 2d geometry of the setup 
\begin{align}\label{eq:diff}
\partial_t m(\varepsilon,\vb{r},t)= D \grad^2  m -\frac{m}{\tau_{\text{rel}}}, 
\end{align}
with magnon density $m(\varepsilon,\vb{r},t)$ per volume and energy, diffusion constant $D$, and spin relaxation time $\tau_\text{rel}$. Here, we assume besides Gilbert damping also elastic, spin-conserving scattering events with a constant mean free path $\ell_\text{mfp}$, leading to an energy-independent diffusion coefficient $D\sim c \,\ell_\text{mfp}$.
Now, in order to keep track of the polarization dynamics, we combine Eq.~\eqref{eq:pol} and Eq.~\eqref{eq:diff} to obtain 
\begin{align}\label{eq:mastereq}
\partial_t{\vb{q}}(\varepsilon,\vb{r},t)=D \grad^2 \vb{q} -\frac{\vb{q}}{\tau_{\text{rel}}}-\boldsymbol{\detuning}(\varepsilon)\times \vb{q} +{\vb{Q}(\varepsilon,{\vb{r}}}),
\end{align} 
where $\vb{q}(\varepsilon,\vb{r},t)$ is the coarse-grained product of the magnon density $m(\varepsilon,\vb{r},t)$ and the associated Stokes vector $\vb{S}(\varepsilon,\vb{r},t)$ describing the polarization. Here, we approximate the dispersion as linear $\varepsilon= \hbar c k$ and write $\boldsymbol{\detuning}(\varepsilon)\equiv\boldsymbol{\detuning}(k)$. For consistency, we introduce a lower cutoff $\Delta<\varepsilon$ with $\Delta\sim \Delta_{x,y}$ anticipating that modes near the gap are of minor importance in spin transport due to their low velocity and high detuning, see Fig.~3.  We remark that Eq.~\eqref{eq:mastereq} is formally equivalent to the pseudospin dynamics in Ref.~\cite{wimmer_2020}, albeit with a classical interpretation using polarization. 

Finally, we added a source term 
\begin{align}
\vb{Q}(\varepsilon,{\vb{r}})=j_{s}^\text{inj}(\varepsilon) \frac{W}{d} \delta(z) \Theta\left(\frac{L}{2}-\vert y \vert\right) \vb{e}_3
\end{align}
which continuously injects a spin current $I^\text{inj}_{s}=\int\mathrm{d}\varepsilon \,j^\text{inj}_{s}(\varepsilon) W L$ at $z=0$ of right circular polarized magnons at the interface. Here, $W$ and $L$ are the width and length of the injector and $d$  is the thickness of the hematite film. As a next step, we estimate the energy dependence of the injected spin current $j_{s}^\text{inj}(\varepsilon)$.

Solving the equation for the spin density $\rho_z=\hbar \vb{e}_3\cdot \vb{q}$ we arrive at the stationary solution
\begin{align}\label{eq:spindens}
\rho_z(\varepsilon,\vb{r})&=\hbar \frac{W}{d} \int\limits_{-\frac{L}{2}}^{\frac{L}{2}} \mathrm{d} y^\prime\int\limits_0^\infty \mathrm{d}\tau G(y-y^\prime,z,\tau) \cos\left[\detuning(\varepsilon)\tau\right]j_{s}^\text{inj}(\varepsilon)\\
&=\hbar \frac{WL}{d}\int\limits_0^\infty  \mathrm{d}\tau g(z,\tau) \cos\left[\detuning(\varepsilon)\tau\right]j_{s}^\text{inj}(\varepsilon). \label{eq:1ddiff}
\end{align}
where we defined the retarded Greens function of the diffusion equation
\begin{align}
G(y,z,\tau)=\Theta(\tau)\frac{1}{4\pi D\tau}e^{-\frac{z^2+y^2}{4D\tau}} e^{-\tau/\tau_\text{rel}}.
\end{align}
Equation~\eqref{eq:spindens} has a very intuitive interpretation. Each magnon with energy $\varepsilon$ that is injected at the source $\vb{r^\prime}=(y^\prime,0)$ propagates diffusively to $\vb{r}$ in a time $\tau$ and changes its polarization periodically with $\detuning(\varepsilon)$.
In the second line of Eq.~\eqref{eq:1ddiff}, we introduced the function 
\begin{align}
g(z,\tau)=\int_{-L/2}^{L/2}\mathrm{d}y\int_{-L/2}^{L/2}\mathrm{d}{y^\prime}G(y-y^\prime,z,\tau)/L,
\end{align}
which for $L\gg z$  effectively reduces to the Greens function of a 1d diffusion equation.

\section{Interfacial spin current}
At the interface, itinerant electrons in the platinum electrode inelastically flip their spin $\hbar/2$ to create a magnon (circular polarized Néel mode) in the antiferromagnet and vice versa, see Fig.~3.
To model this interfacial spin current, we use the heuristic formula
\begin{align}
j^\text{IF}_s(\varepsilon)=  V\frac{\hbar g^{\uparrow\downarrow}_n}{4\pi} \vb{e}_z\cdot{\ev{\vb{n}\times \vb{\dot{n}}} }{\cal N}(\varepsilon)\label{eq:spininj1}
\end{align}
describing that the injected spin with energy $\varepsilon$ induces a circular polarized Néel mode $\vb{n}=\Re\left(\vb{n}_0+\delta \vb{n}\right)$ given by
\begin{align}
    \delta \vb{n}=C \vb{p}_\circlearrowleft e^{-i\varepsilon t/\hbar}=\frac{C}{\sqrt{2}} (\vb{e}_x+i\vb{e}_y) e^{-i\varepsilon t/\hbar},
\end{align}
where the frequency is given by $\varepsilon/\hbar$. $g^{\uparrow\downarrow}_n$ is the interfacial spin-mixing conductance per area and the time average $\ev{\ldots}=\frac{\varepsilon}{2\pi \hbar} \int_0^{2\pi \hbar/\varepsilon}\mathrm{d}t\ldots$ is performed over one period. 
The density of states is given by ${\cal N}(\varepsilon)=\frac{\varepsilon^2}{2\pi^2c^3 \hbar^2}$, where we approximate the dispersion as linear (gray dashed line), $\varepsilon=\hbar c k$, see Fig.~3. Modes near the gap will be ignored due to their slow velocity and high detuning.
We can rewrite Eq.~\eqref{eq:spininj1} in terms of the solid angle $\Omega=\pi C^2/2$ enclosed by the Néel order precession. We find
\begin{align}
j^\text{IF}_s(\varepsilon)=  \frac{ g^{\uparrow\downarrow}_n}{4\pi^2}{\varepsilon\,\Omega(\varepsilon)}{\cal N}(\varepsilon).\label{eq:spininj2}
\end{align}
In the spirit of Bohr quantization, this solid angle can be decomposed as
\begin{align}
    \Omega(\varepsilon)=n^\text{IF}\Omega_1(\varepsilon),
\end{align}
i.e., it is quantized in units of the solid angle $\Omega_1$ associated with a single magnon.
The latter can be determined by the condition that $\rho_z V=\hbar$ if a single magnon is absorbed. Using Eq.~\eqref{eq:spinvspolarization} for $\tau_\text{rel}\rightarrow \infty$ and $S_3=+1$, we obtain
\begin{align}
  \rho_z V = \frac{V \chi}{\hbar \pi} \varepsilon\,{\Omega_1}(\varepsilon)=\hbar \quad \Rightarrow \quad  \Omega_1(\varepsilon)=\frac{\hbar^2\pi}{\chi V} \frac{1}{\varepsilon}.
\end{align}  
Finally, we assume that the number $n^\text{IF}$ is given by the imbalance
\begin{align}
    n^\text{IF}=n_\text{BE}(\varepsilon-\mu)-n(\varepsilon),
\end{align}
where $n_\text{BE}(\varepsilon-\mu)$ describes the spin-flip excitations in the platinum electrode\footnote{In a metal the underlying particles are electrons (fermions). Nevertheless, a Bose-Einstein distribution appears via the relation $n_\text{BE}(\varepsilon-\mu)=(\varepsilon-\mu)^{-1}\int \mathrm{d} E \,n_\text{FD}(E+\varepsilon-\mu_\uparrow)\left[1- n_\text{FD}(E-\mu_\downarrow)\right]$, where $n_\text{FD}(\varepsilon)=(e^{\varepsilon/k_\text{B}T}+1)^{-1}$ is the Fermi-Dirac distribution with electrochemical potentials $\mu_{\uparrow,\downarrow}$ and the spin accumulation is given by $\mu=\mu_\uparrow-\mu_\downarrow$. Thus, the spin-flip excitations (absorption of spin-up electron and emission of spin-down electron) effectively show similarities to bosonic degrees of freedom.} and $n(\varepsilon)$ the number of magnons in the antiferromagnet. Here, $n_\text{BE}(\epsilon)=(e^{ \epsilon/k_\text{B}T}-1)^{-1}$ is the Bose-Einstein distribution and $\mu$ describes the spin accumulation. 
Putting everything together, the energy-dependent interfacial spin current becomes 
\begin{align}
j^\text{IF}_s(\varepsilon)&=\frac{\hbar^2 g^{\uparrow\downarrow}_n}{4\pi \chi}  {\cal N}(\varepsilon)\left[n_\text{BE}(\varepsilon-\mu)-n(\varepsilon)\right]. 
\end{align}
Interestingly, the energy dependence of $\varepsilon \,\Omega_1(\varepsilon)$ has cancelled.  
Now, we can evaluate both the spin injection and detection. 

\subsection{Spin injection}
At the injector [$\vb{r}=(y,0)$ with $-L/2<y<L/2$], spins are entering from the platinum electrode due to the spin accumulation $\mu>0$. For  the antiferromagnet, we use $n(\varepsilon)\approx n_\text{BE}(\varepsilon)$, assuming the interface acts as a bottleneck and the backflow of magnons into the platinum electrode can be ignored. We obtain
\begin{align}
j^\text{inj}_s(\varepsilon)&\approx\frac{\hbar^2 g^{\uparrow\downarrow}_n}{4\pi \chi}  {\cal N}(\varepsilon)\left[n_\text{BE}(\varepsilon-\mu)-n_\text{BE}(\varepsilon)\right] \\
&\approx \frac{\hbar^2 g^{\uparrow\downarrow}_n}{4\pi \chi}  {\cal N}(\varepsilon)\left[- \partial_{ \varepsilon} n_\text{BE}(\varepsilon) \mu\right] \\
&\approx \frac{g^{\uparrow\downarrow}_n}{(2\pi c)^3\hbar\chi} \mu \,k_\text{B}T,
\end{align}
where, in the second line, we expand in $\mu$ and, in the third line, we assume the Rayleigh-Jeans limit ($\varepsilon \ll k_\text{B}T$) and insert the density of states. Remarkably, the current becomes completely energy independent.

\subsection{Spin detection}
For the detector [$\vb{r}=(y,0)$ with $-L/2<y<L/2$], the platinum electrode is in thermal equilbrium with $\mu=0$. Now, the excess magnons are on the antiferromagnetic side and we write $n(\varepsilon)=n_\text{BE}(\varepsilon) + \Delta n(\varepsilon,\vb{r})$.
We find 
\begin{align}
   j^\text{det}_s(\varepsilon,\vb{r})& = \frac{\hbar^2 g^{\uparrow\downarrow}_n}{4\pi\chi} {\cal N}(\varepsilon)\Delta n(\varepsilon,\vb{r})\\
   &=  \frac{\hbar g^{\uparrow\downarrow}_n}{4\pi\chi}  \rho_z(\varepsilon,\vb{r}).
\end{align}
Here we use that the number of excess circular polarized magnons $\Delta n$ is given by the spin density $\rho_z=\hbar \Delta n\,{\cal N}(\varepsilon)$. We remark that   $j^\text{det}_s\sim(g^{\uparrow\downarrow}_n)^2$, which, retrospectively, allowed us to omit the drain term (the counterpart to $\vb{Q}$ at the detector) in the magnon kinetic equation.

Finally, in the detector, this spin current $j^\text{det}_s$ in turn induces a charge current $j^\text{det}_c$ via the inverse spin Hall effect, $j^\text{det}_c=\theta j^\text{det}_s$. Using Ohms law to calculate the resulting electric field, ${\sigma}^{-1} j_c^\text{det}$, the detected voltage is given by 
\begin{align}
    U_\text{det}&=\sigma^{-1}\theta\int\limits_{-L/2}^{L/2} \mathrm{d}y \int \mathrm{d}\varepsilon \,j^\text{det}_s(\varepsilon,y,\ell)\\ \label{eq:udet}
    &\approx \frac{ \theta }{\sigma}\frac{ \hbar^2 g^{\uparrow\downarrow}_n}{4\pi \chi} \frac{WL}{d} j_s^\text{inj}\int \limits_{\Delta}^{k_\text{B}T} \mathrm{d}\varepsilon \int\limits_0^\infty \mathrm{d}\tau g(\ell,\tau) \cos\left[\detuning(\varepsilon) \tau\right],
\end{align}
where $\theta$ is proportional to the spin Hall angle~\cite{keller_2018} and $\sigma$ is the electrical conductivity. In the second line, we roughly limit the integration to $\Delta<\varepsilon<k_\text{B}T$, see Fig.~3. The upper limit allows us to use the Rayleigh-Jeans limit, while the lower limit $\Delta\sim \Delta_{x,y}$ ignores states near the band gap due to their slow velocity and large detuning.

\bibliography{References}
